\documentclass{iau} 

\usepackage{graphicx}
\usepackage{caption}
\usepackage{siunitx} 
\usepackage{graphicx}
\usepackage{longtable}
\usepackage{float}

\newcommand{\tablenote}[1]{\parbox{18.3cm}{\indent \footnotesize{#1}}}

\DeclareCaptionFormat{cont}{#1 (cont.)#2#3\par} % continued figures

\title[SiC$_2$ in Carbon Star Envelopes~~ ] %% give here short title %%
{The Abundance of SiC$_2$ \\ in Carbon Star Envelopes}

\author[]   %% give here short author list %%
{S.~Massalkhi$^1$, M.~Ag\'undez$^1$, J.~Cernicharo$^1$, \\ J.P. Fonfr\'ia$^1$, and M. Santander-Garc\'ia$^1$}

\affiliation{$^1$Grupo de Astrof\'isica Molecular, Instituto de Ciencia de Materiales de Madrid, CSIC \\ C/ Sor Juana In\'es de la Cruz 3, 28049 Cantoblanco, Spain \\ }

\pubyear{2017}
\volume{IAUS 332}  %% insert here IAU Symposium No.
\jname{Astrochemistry VII – Through the Cosmos from Galaxies to Planets}
\editors{M. Cunningham, T. Millar \& Y. Aikawa, eds.}

\begin{document}

\maketitle

\begin{abstract}
Silicon carbide dust grains are ubiquitous in circumstellar envelopes around C-rich AGB stars. However, the main gas-phase precursors leading to the formation of SiC dust have not yet been identified. To date, only three molecules containing an Si--C bond have been identified to have significant abundances in C-rich AGB stars: SiC$_2$, SiC, and Si$_2$C. The ring molecule SiC$_2$ has been observed in a handful of evolved stars, while SiC and Si$_2$C have only been detected in the C-star envelope IRC\,+10216. We aim to study how widespread and abundant SiC$_2$, SiC, and Si$_2$C are in envelopes around C-rich AGB stars and whether or not these species play an active role as gas-phase precursors of silicon carbide dust in the ejecta of carbon stars.
%\keywords{..., ..., ..., ...}
\end{abstract}
\firstsection % if your document starts with a section,
\section{Setting the Scene}
During the late stages of their evolution, asymptotic giant branch (AGB) stars experience significant mass loss processes, which result in extended circumstellar envelopes (CSEs). These environments are efficient factories of molecules and dust grains. The main paradigm for the dust formation process involves a first step in which condensation nuclei of nanometer size are formed from some gas-phase precursors of highly refractory character and a second step in which the nuclei grow to micrometer sizes by accretion and coagulation as the material is pushed out by the stellar wind. The chemical nature of the molecules and dust grains formed depends to a large extent on the C/O elemental abundance ratio at the stellar surface. Although much has been advanced recently, there is still much to understand about how dust grains are formed and which are the main gas-phase precusors. This is the main driver of the ERC Synergy Project NANOCOSMOS.

\begin{table}
\caption{Sample of carbon stars}\label{table:sources}
\centering
\begin{tabular}{lccccccc}
\hline \hline
\multicolumn{1}{l}{Name} & \multicolumn{1}{c}{$V_{\rm LSR}$} & \multicolumn{1}{c}{$D$} & \multicolumn{1}{c}{$T_{\star}$} & \multicolumn{1}{c}{$L_{\star}$} & \multicolumn{1}{c}{$\dot{M}$} & \multicolumn{1}{c}{$V_{\rm exp}$} & \multicolumn{1}{c}{$f_0$}\\

\multicolumn{1}{c}{}  & \multicolumn{1}{c}{(km~s$^{-1}$)} & \multicolumn{1}{c}{(pc)} & \multicolumn{1}{c}{(K)} & \multicolumn{1}{c}{(L$_{\odot}$)}  & \multicolumn{1}{c}{(M$_{\odot}$ yr$^{-1}$)} & \multicolumn{1}{c}{(km~s$^{-1}$)} & \\
\hline
IRC\,+10216 & $-26.5$ & 130 $^{[a]}$  & 2330 $^{[a]}$  &  8750 $^{[a]}$  & $2.0\times10^{-5}$ $^{[a]}$  & 14.5  & $3.7\times10^{-7}$ \\
CIT\,6    & $-1$    & 440 $^{[g]}$ & 1800 $^{[g]}$ & 10000 $^{[g]}$ & $6.0\times10^{-6}$ $^{[g]}$  & 17 & $1.1\times10^{-5}$  \\
CRL\,3068 & $-31.5$  & 1300 $^{[g]}$ & 1800 $^{[g]}$  & 10 900 $^{[g]}$ & $2.5\times10^{-5}$ $^{[g]}$  & 14.5  & $1.9\times10^{-6}$\\ 
S\,Cep  & $-15.3$     & 380 $^{[k]}$    & 2200 $^{[k]}$   & 7300 $^{[k]}$  & $1.2\times10^{-6}$ $^{[k]}$  & 22.5  & $1.0\times10^{-5}$ \\
IRC\,+30374 & $-12.5$     & 1200 $^{[j]}$   & 2000 $^{[j]}$   & 9800 $^{[j]}$   & $1.0\times10^{-5}$ $^{[j]}$     & 25 & $9.7\times10^{-6}$\\
Y\,CVn & $+22$       & 220 $^{[k]}$    & 2200 $^{[k]}$   & 4400 $^{[k]}$   & $1.5\times10^{-7}$ $^{[k]}$  & 7  & $4.0\times10^{-6}$\\
LP\,And & $-17$       & 630 $^{[g]}$    & 1900 $^{[g]}$   & 9600 $^{[g]}$  & $7.0\times10^{-6}$ $^{[g]}$ &14.5   &$6.8\times10^{-6}$\\
V\,Cyg  & $+13.5$     & 366 $^{[g]}$    & 2300 $^{[g]}$   & 6000 $^{[g]}$  & $1.6\times10^{-6}$ $^{[g]}$   & 12 & $6.6\times10^{-6}$\\	
UU\,Aur  & $+6.7$  & 260 $^{[k]}$   & 2800 $^{[k]}$   & 6900 $^{[k]}$ & $2.4\times10^{-7}$ $^{[k]}$   & 10.6 & $<1.0\times10^{-6}$\\ 
V384\,Per & $-16.8$     & 560 $^{[k]}$   & 2000 $^{[k]}$   & 8100 $^{[k]}$   & $2.3\times10^{-6}$ $^{[b]}$  & 15.5  & $1.3\times10^{-5}$\\
IRC\,+60144 & $-48.8$     & 1030 $^{[b]}$  & 2000 $^{[b]}$ & 7800 $^{[b]}$   & $3.7\times10^{-6}$ $^{[b]}$   & 19.5  & $6.0\times10^{-6}$\\
U\,Cam & $+6$        & 430 $^{[i]}$    & 2695 $^{[i]}$  & 7000 $^{[i]}$ & $2.0\times10^{-7}$ $^{[i]}$  & 13 & $3.7\times10^{-5}$\\ 
V636\,Mon   & $+10$       & 880 $^{[f]}$   & 2500 $^{[n]}$   & 8472 $^{[e]}$  & $5.8\times10^{-6}$ $^{[f]}$    & 20   & $1.7\times10^{-6}$\\	
IRC\,+20370  & $-0.8$      & 600 $^{[k]}$    & 2200 $^{[k]}$   & 7900 $^{[k]}$   & $3.0\times10^{-6}$  $^{[b]}$     & 14  & $4.2\times10^{-6}$\\
R\,Lep  & $+11.5$     & 432 $^{[b]}$   & 2200 $^{[b]}$  & 5500 $^{[b]}$  & $8.7\times10^{-7}$ $^{[b]}$   & 17.5 & $<2.7\times10^{-6}$\\		
W\,Ori   & $-1$        & 220 $^{[k]}$    & 2600 $^{[k]}$   & 3500 $^{[k]}$   & $7.0\times10^{-8}$ $^{[k]}$     & 11  & $1.4\times10^{-5}$\\
CRL\,67 & $-27.5$     & 1410 $^{[d]}$    & 2500 $^{[n]}$  & 9817 $^{[d]}$    & $1.1\times10^{-5}$ $^{[d]}$    & 16&   $1.0\times10^{-5}$\\
CRL\,190 & $-39.5$     & 2790 $^{[d]}$    & 2500 $^{[c]}$  & 16 750 $^{[d]}$   & $6.4\times10^{-5}$ $^{[d]}$      & 17.0   & $8.8\times10^{-7}$\\
S\,Aur  & $-17$       & 300 $^{[k]}$    & 3000 $^{[k]}$   & 8900 $^{[k]}$   & $4.0\times10^{-7}$ $^{[k]}$   & 24.5   & $7.3\times10^{-6}$\\ 
V\,Aql  & $+53.5$     & 330 $^{[k]}$    & 2800 $^{[k]}$   & 6500 $^{[k]}$   & $1.4\times10^{-7}$ $^{[k]}$   & 8 & $2.0\times10^{-5}$\\
CRL\,2513   & $+17.5$     & 1760 $^{[d]}$   & 2500  $^{[c]}$  & 8300 $^{[c]}$ & $2.0\times10^{-5}$ $^{[d]}$    & 25.5  & $1.6\times10^{-6}$\\
CRL\,2477 & $+5$        & 3380 $^{[m]}$   & 3000 $^{[m]}$ & 13 200 $^{[m]}$  & $1.1\times10^{-4}$ $^{[m]}$    & 20  & $6.0\times10^{-7}$\\
CRL\,2494  & $+29$       & 1480 $^{[b]}$    & 2400 $^{[b]}$    & 10 200 $^{[b]}$  & $7.5\times10^{-6}$ $^{[b]}$   & 20  & $2.7\times10^{-5}$\\
RV\,Aqr & $+0.5$      & 670 $^{[k]}$    & 2200 $^{[k]}$     & 6800 $^{[k]}$     & $2.3\times10^{-6}$ $^{[b]}$   & 15 & $3.0\times10^{-6}$\\
ST\,Cam  &  -13.6     & 360 $^{[k]}$   & 2800 $^{[k]}$    & 4400 $^{[k]}$   &  $1.3\times10^{-7}$ $^{[k]}$   & 8.9 & $<4.0\times10^{-6}$\\ 
\hline
\end{tabular}
\tablenote{\\ The values $f_{0}$ in the last column are the derived SiC$_{2}$ abundances relative to H$_{2}$.\\
References: $^{[a]}$~Ag\'undez~et~al.~(2012), $^{[b]}$~Danilovich~et~al.~(2015), $^{[c]}$~Groenewegen~et~al.~(1998), \\ $^{[d]}$~Groenewegen~et~al.~(2002), $^{[e]}$~Guandalini~et~al.~(2013), $^{[f]}$~Guandalini~et~al.~(2006), \\ $^{[g]}$~Ramstedt~et~al.~(2014), $^{[i]}$~parameters of the present-day wind of U\,Cam from Sch\"oier et al.~(2005),\\ $^{[j]}$~Sch\"oier et al.~(2006), $^{[k]}$~Sch\"oier et al.~(2013), $^{[m]}$~Speck~et~al.~(2009), \\ $^{[n]}$ Assumed value for the stellar effective temperature  $T_{\star}$ is 2500 K}
\end{table}

\section{SiC Dust: What is it about?}
Silicon carbide (SiC) dust grains, which are detected through a solid-state band at \SI{11.3}{\micro\metre}, are exclusively found in the envelopes around C-type stars (C/O $>$ 1). Here, we explore which are the main precursors of SiC dust grains. Only three gas-phase molecules containing the Si-C bond have been observed with significant abundances in C-rich envelopes around AGB stars. The ring molecule SiC$_2$ has been observed towards a few AGB and post-AGB stars (\cite[Thaddeus et al. 1984]{Thaddeus84}; \cite[Bachiller et al. 1997]{Bachiller97}; \cite[Zhang et al. 2009a]{Zhang2009a},\cite[b]{Zhang2009b}), while SiC and Si$_2$C have been only observed in the C star envelope IRC\,+10216 (\cite[Cernicharo et al. 1989]{Cernicharo89}, \cite[2015b]{Cernicharo15}). In addition to these species, other Si-bearing species detected in IRC\,+10216 are SiCN (\cite[Gu\'elin et al. 2000]{Guelin2000}), SiNC (\cite[Gu\'elin et al. 2004]{Guelin2004}), SiC$_3$ (\cite[Apponi et al. 1999]{Apponi99}) and SiC$_4$ (\cite[Ohishi et al. 1989]{Ohi89}) but the observed line profiles suggest that they are formed in the external layers of the envelope through the reactions of C$_2$H$_2$ and SiC$_2$ and their reaction products.

\begin{figure}
\centering
\includegraphics[scale=0.4]{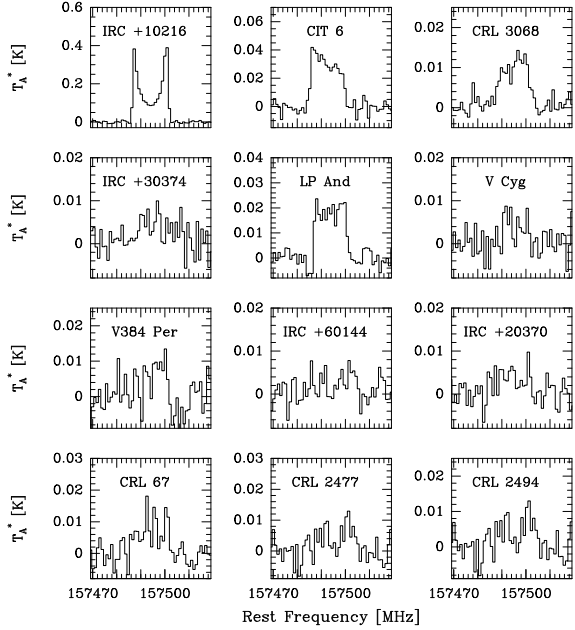}
\caption{SiC $^3\Pi_2$ $J=4-3$ line observed in 12 out of the 25 sources of our sample. The spectral resolution is 1 MHz.}
\label{fig:sic_emission_lines}
\end{figure}

Much of the knowledge about the role of the molecules SiC$_2$, SiC, and Si$_2$C as precursors of SiC dust grains comes from the study of IRC\,+10216, as in this source SiC$_2$ has been thoroughly identified across the mm and sub-mm ranges with ground based radio telescopes and with the Herschel Space Telescope (\cite[Lucas et al. 1995]{Lucas95}; \cite[Cernicharo et al. 2010]{Cernicharo2010}; \cite[M\"uller et al. 2012]{Muller2012}, \cite[Fonfr\'ia et al. 2014]{Fonfria14}; \cite[Velilla Prieto et al. 2015]{Velilla15}). The paradigm emerged from the studies of IRC\,+10216 suggests that only SiC$_2$ and Si$_2$C are present in the innermost regions, while SiC is probably a photodissociation product of these molecules, and it is thus restricted to the outer envelope. The U-shaped line profiles of SiC (\cite[Cernicharo et al. 2000]{Cernicharo2000}) and the upper limits derived to its abundance (\cite[Velilla Prieto et al. 2015]{Velilla15}) in the inner envelope support this idea. Moreover, chemical equilibrium calculations predict abundant SiC$_2$ and Si$_2$C, and little SiC in the hot and dense surroundings of the AGB star (\cite[Tejero \& Cernicharo 1991]{Tej1991}; \cite[Yasuda \& Kozasa 2012]{Yasuda12}; \cite[Cernicharo et al. 2015b]{Cernicharo15}). 
It therefore seems that the gas-phase molecule SiC is not an important building block of SiC dust grains, while SiC$_2$ and Si$_2$C are probably the main gas-phase precursors.
\begin{table}
\caption{Rotational transitions of SiC$_{2}$, Si$_2$C, and SiC covered}\label{table:lines}
\centering
\scalebox{0.9}{
\begin{tabular}{lccr}
\hline
\multicolumn{1}{l}{Transition}  & \multicolumn{1}{c}{Frequency} & \multicolumn{1}{c}{$A_{ul}$} & \multicolumn{1}{c}{$E_u$} \\
\multicolumn{1}{l}{}                 & \multicolumn{1}{c}{(MHz)}  & \multicolumn{1}{c}{(s$^{-1}$)} & \multicolumn{1}{c}{(K)} \\
\hline
\multicolumn{4}{c}{SiC$_2$} \\
\hline
$6_{2,5}-5_{2,4}$ & 140920.171 & 7.65 $\times$ 10$^{-5}$ & 31.5 \\
$6_{4,3}-5_{4,2}$ & 141751.492 & 4.87 $\times$ 10$^{-5}$ & 55.0 \\
$6_{4,2}-5_{4,1}$ & 141755.360 & 4.87 $\times$ 10$^{-5}$ & 55.0 \\
$6_{2,4}-5_{2,3}$ & 145325.875 & 8.39 $\times$ 10$^{-5}$ & 32.0 \\
$7_{0,7}-6_{0,6}$ & 158499.228 & 1.23 $\times$ 10$^{-4}$ & 31.0 \\
\hline
\multicolumn{4}{c}{Si$_2$C} \\
\hline
$11_{1,11}-10_{0,10}$ & 144033.475 & 9.57 $\times$ 10$^{-6}$ & 29.3 \\
$22_{2,20}-22_{1,21}$ & 155600.100 & 1.54 $\times$ 10$^{-5}$ & 115.0 \\
$13_{1,13}-12_{0,12}$ & 157768.156 & 1.28 $\times$ 10$^{-5}$ & 39.3 \\
$20_{2,18}-20_{1,19}$ & 157959.754 & 1.54 $\times$ 10$^{-5}$ & 97.4 \\
$18_{2,16}-18_{1,17}$ & 160644.790 & 1.55 $\times$ 10$^{-5}$ & 81.4 \\
\hline
\multicolumn{4}{c}{SiC} \\
\hline
$^3\Pi_2$ $J=4-3$ & 157494.101 & 3.98 $\times$ 10$^{-5}$ & 13.2 \\
\hline
\end{tabular}
}
\end{table}
However, little is still known on how widespread these molecules are in other carbon stars, what are their relative abundances, and what is their role on the formation of SiC dust. To explore the role of gas-phase SiC$_2$ molecules on the formation of silicon carbide dust, we carried out sensitive observations with the IRAM 30m telescope of a sample of 25 C-rich AGB stars (see Table \ref{table:sources}) to observe emission lines of SiC$_2$, Si$_2$C and SiC in the $\lambda$ 2mm band (see Table \ref{table:lines}). The observations resulted in the detection of SiC$_2$ in 22 out of the 25 targeted sources. The lines corresponding to the rotational transitions $6_{4,3}-5_{4,2}$ and $6_{4,2}-5_{4,1}$ appear blended although in most sources each of these lines could be fitted individually. We also report the detection of SiC in 12 of the 25 targeted C-rich AGB envelopes (see Fig.\ref{fig:sic_emission_lines}). We note that prior to this study, this radical had been detected only in the C-star envelope IRC\,+10216 (\cite[Cernicharo et al. 1989]{Cernicharo89}). The molecule Si$_2$C was not detected in any of the targeted sources, at the exception of IRC\,+10216, which at present is the only source in which this molecule has been identified (\cite[Cernicharo et al. 2015b]{Cernicharo15}). \\
Hereafter, we focus on the analysis of SiC$_2$ and leave the interpretation of other molecules for future studies. We aim to model its emission and determine its abundance in each source to provide a global view of how abundant this molecule is in circumstellar envelopes around C-rich AGB stars. 
 
\section{Abundance of SiC$_2$: How we did it}
To model all sources, we adopted a common physical scenario consisting of a spherically symmetric envelope of gas and dust expanding with a constant velocity and mass-loss rate around a central AGB star. The spherical envelope is described by the radial profile of various physical quantities, such as the gas density, the kinetic temperature of the gas, the dust temperature, the expansion velocity, and the microturbulence velocity. We also consider a dusty component of the envelope, although, we noticed that dust plays a minor role in the excitation of the rotational lines of SiC$_2$ observed.

We performed non-LTE excitation and radiative transfer calculations to model the line emission of SiC$_{2}$ based on the multi-shell LVG (Large Velocity Gradient) method of the MADEX code (\cite[Cernicharo, J. 2012]{Cernicharo12}). The circumstellar envelope is divided into a number of concentric shells, each of them with a characteristic set of physical properties and SiC$_{2}$ abundance, and statistical equilibrium equations are solved in each of them. In each shell, the contribution of the background radiation field (cosmic microwave background, stellar radiation, and thermal emission from surrounding dust) is included. In the excitation calculations, we consider rotational levels up to $J=39$ (i.e., a total number of 620 energy levels) within the ground vibrational state of SiC$_2$. The rate coefficients through inelastic collisions with H$_2$ and He were taken from (\cite[Chandra, S. \& Kegel, W. H. 2000]{Chanra2000}). We consider that SiC$_2$ is formed close to the star with a given fractional abundance, which remains constant across the envelope up to the outer regions, where it is photodissociated by the ambient ultraviolet radiation field of the local Interstellar Medium. To model the emission lines of SiC$_2$ and determine its abundance, we varied the initial fractional abundance of SiC$_2$ relative to H$_2$, $f_0$, until the calculated line profiles matched the observed ones. We choose as best-fit model that which results in the best overall agreement between calculated and observed line profiles for the entire set of SiC$_2$ lines observed. In those cases in which no lines of SiC$_2$ are detected, we derive upper limits to the abundance of SiC$_2$ by choosing the maximum abundance that results in line intensities compatible with the noise level of the observations.

\section{Results: What We Found}
The calculated line profiles resulting from our best-fit model for each of the sources is compared with the observed line profiles in Fig.~\ref{fig:sic2_lines}. In most sources, the shapes of the SiC$_2$ lines observed are nearly flat-topped, which is indicative of optically thin emission not resolved by the 15.0-17.5$''$ beam of the IRAM 30m telescope. One notable exception is IRC\,+10216, whose close proximity (130 pc) makes the emission to be spatially resolved by the telescope beam and line profiles show a marked double-peaked shape.

\begin{figure*}
\centering
\includegraphics[scale=0.3]{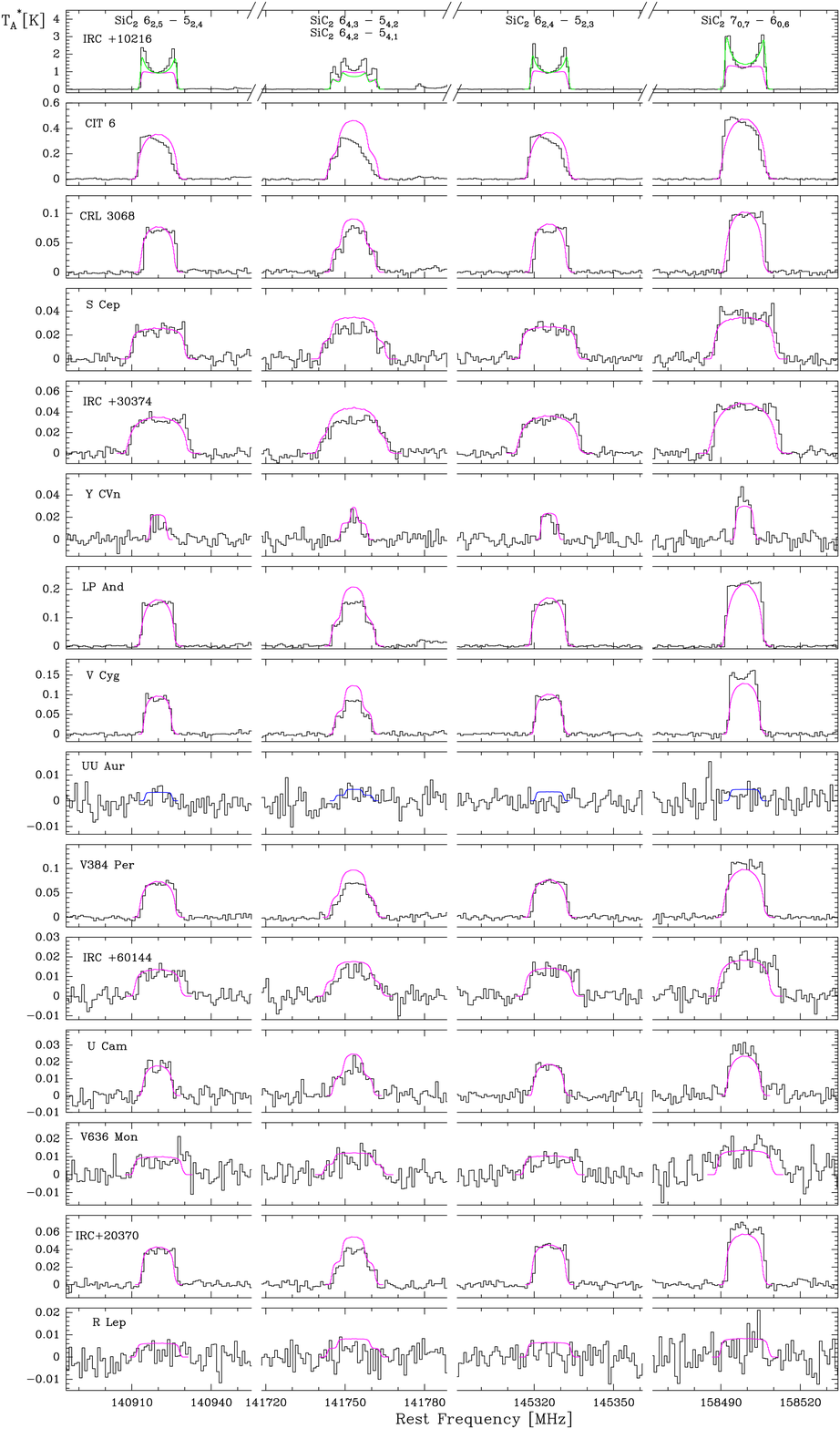}
\caption{The black histograms are the rotational lines of SiC$_{2}$ observed with the IRAM 30m telescope in the 25 target sources. The spectral resolution of the observed lines is 1 MHz. The magenta lines are the calculated emerging line profiles from the best-fit model using the LVG method. The blue lines are the predicted line profiles in sources in which SiC$_{2}$ was not detected. In the panel of IRC +10216, the green lines correspond to the lines profiles calculated assuming LTE.}
\label{fig:sic2_lines}
\end{figure*}

\begin{figure*}
\ContinuedFloat
\captionsetup{list=off,format=cont}
\centering
\includegraphics[scale=0.3]{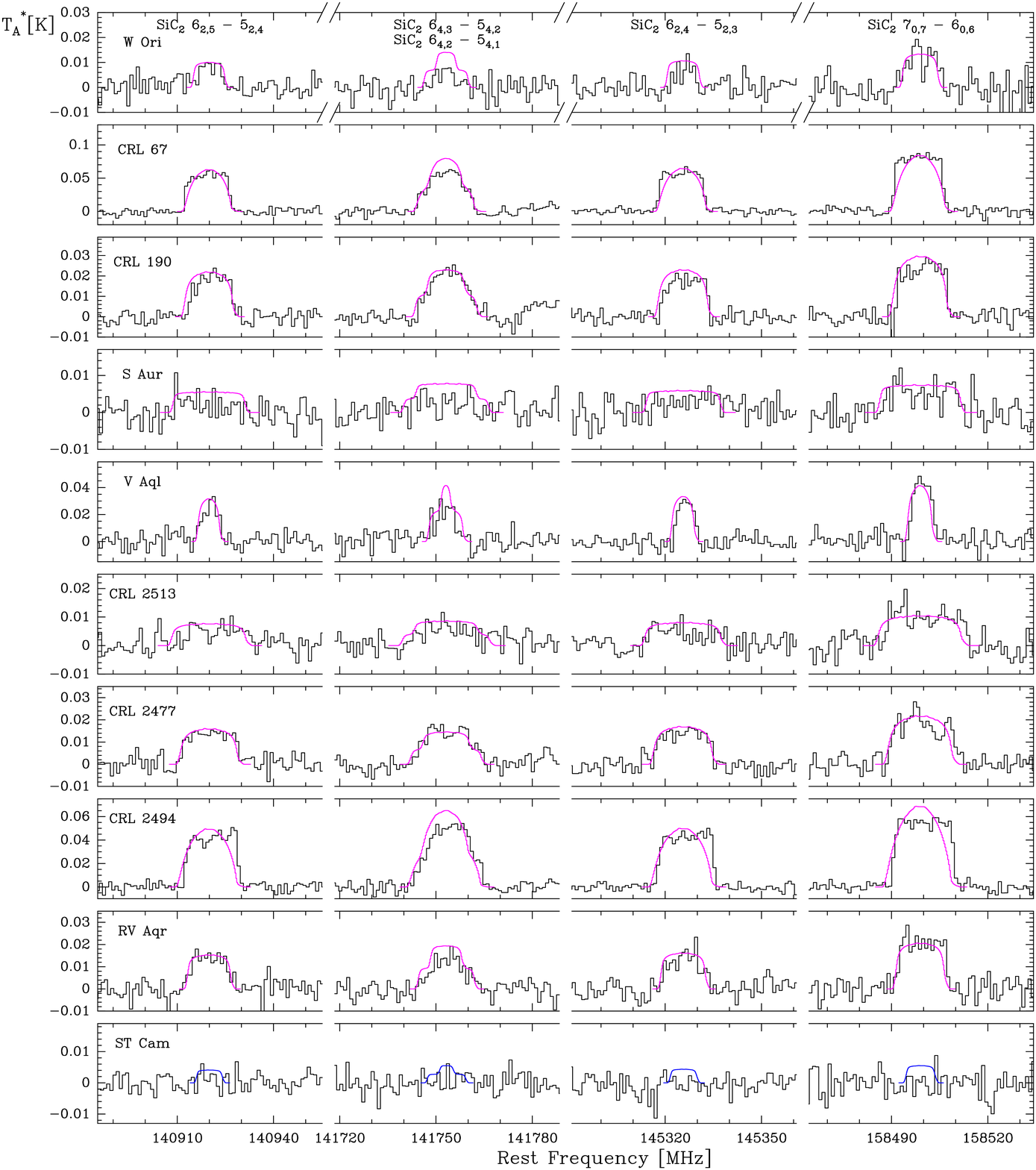}
\caption{}
\end{figure*}

\begin{figure}[h]
\begin{center}
 \includegraphics[scale=0.16]{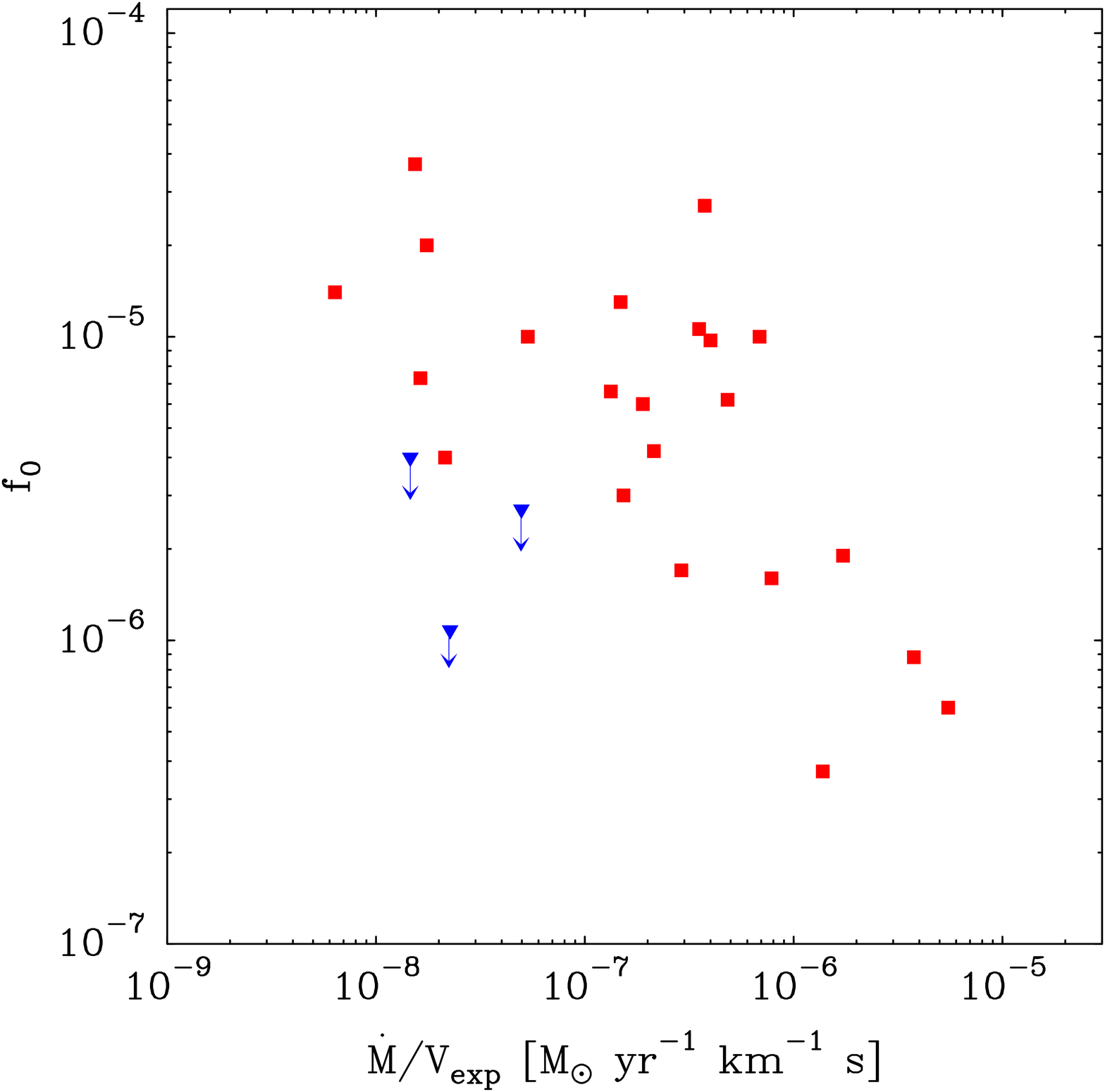} 
% \vspace*{-1.0 cm}
 \caption{SiC$_2$ fractional abundance $f_0$ obtained from radiative transfer analysis, as a function of a density measure ($\dot{M}/V_{exp}$) for the C-rich AGB stars are represented with filled square. Upper limits for the non-detections are denoted with downward triangles.}
   \label{fig:abun}
\end{center}
\end{figure}

The fractional abundance of SiC$_2$ shows an interesting trend with the density in the envelope, evaluated through the quantity $\dot{M}/V_{\rm exp}$. As shown in Fig.~\ref{fig:abun}, SiC$_2$ becomes less abundant as the density in the envelope increases. In a standard scenario describing the chemistry of an expanding envelope around an AGB star, the abundance with which a molecule is injected into the intermediate and outer envelope is set by the processes occurring in the inner regions.  In this sense, the abundance of SiC$_2$ in the intermediate regions of the envelope is set by thermochemical equilibrium (TE) at the stellar surface and is possibly modified by different processes during expansion. We carried out TE calculations which show that the SiC$_2$ abundances are insensitive to the gas density in the region where the molecular abundances quench to the TE values (see further, \cite[Massalkhi et al. 2017]{Massalkhi17}). We interpret this as that the SiC$_2$ molecules deplete from the gas phase to incorporate into solid dust grains, a process that is favored at higher densities owing to the higher rate at which collisions between particles occur. We note that the emission from the five SiC$_2$ lines observed in our sources arises typically from intermediate regions of the envelope, at radial distances in the range 10$^{15}$-10$^{16}$ cm where dust formation has already taken place, and therefore the SiC$_2$ abundances have to be considered as post-condensation abundances. Further support for our scenario comes from mm-wave interferometric observations of SiC$_2$ in the C-star envelope IRC\,+10216 (\cite[Lucas et al. 1995]{Lucas95}; \cite[Fonfr\'ia et al. 2014]{Fonfria14}; \cite[Velilla Prieto et al. 2015]{Velilla15}), which shows that SiC$_2$ is present in regions close to the star, then experiences a marked abundance decline at 10-20 R$_*$ (very likely due to condensation onto dust grains), and appears again in the outer envelope (probably as a result of the interaction between the UV radiation field and the envelope). The ring molecule SiC$_2$ thus emerges as a very likely gas-phase precursor in the process of formation of SiC dust in envelopes around C-rich AGB stars.

\section{Similarly? C$_2$ in AGB stars}
The dependence of the SiC$_2$ gas-phase abundance in the envelopes of different AGB stars on the mass-loss rate is similar to what we have found for C$_2$ in a sample of C-rich AGB stars (\cite[Fonfr\'ia et al. 2017]{Fonfria17}).
The observations of these stars were taken in the optical and near-IR wavelength ranges with the high spectral resolution spectrograph CARMENES ($R\simeq 85,000$) mounted on the 3.5~m telescope at the German-Spanish Astronomical Center at Calar Alto (CAHA).
This project represents a big effort to understand the chemical evolution of C-rich AGB stars in their very complex, warm inner envelopes where the dust is formed and the gas expansion mechanism begins to work.

C$_2$ is a molecule found in C-rich AGB stars some decades ago (e.g., \cite[Bakker et al. 1995]{Bakker95}).
The spectrum of this molecule shows two strong electronic systems in the near-IR and the optical comprising a large number of vibro-electronic bands: the Phillips system ($\textnormal A^1\Pi_u-\textnormal X^1\Sigma_g$ centered at 11900\AA) and the Swan system ($\textnormal d^3\Pi_g-\textnormal a^3\Pi_u$ centered at 5200\AA).
No rotational or ro-vibrational electric dipolar transition is allowed in the ground electronic state of this homonuclear molecule, making its detection at wavelengths longer than $\simeq 2.5~\mu$m pretty difficult.
The analysis of both systems is very useful to estimate the C$_2$ abundance with respect to H$_2$ in gas phase, the physical conditions, and the gas kinematics in the shells of the envelope where C$_2$ exists.
Determining the fraction of C$_2$ in the $\textnormal a^3\Pi_u$ electronic state is of particular interest since the chemical reactivity of this molecule highly depends of the characteristics of its valence electrons, being significantly more reactive if it is electronically excited.

In our data, we have detected both systems.
The obvious detection of the Swan system means that its lower electronic state $\textnormal a^3\Pi_u$, which is at $\simeq 1000$~K, is significantly populated.
This is an indication of warm C$_2$ that can be found only close to the photosphere of the central stars and can participate in the chemistry in high-temperature, high-density environments.
It is believed that C$_2$ arises from carbon atoms recently ejected by the stellar pulsation that rapidly combine between them into carbon clusters (C$_2$, C$_3$, C$_4$,\ldots), something that we are investigating with the Stardust machine, developed in the frame of the ERC project NANOCOSMOS.
To date, it remains unknown if there is an additional contribution to the observed spectra from cold C$_2$ formed in the outer shells of the envelopes due to the active chemistry triggered by the external dissociating radiation field, as it happens with, e.g., C$_3$ and C$_5$ (\cite[Hinkle et al. 1988]{Hinkle88}, Bernath et al. 1989).
However, we expect this to be elucidated from our on-going research, which will allow us to describe the C$_2$ abundance profile throughout the whole envelope taking advantage of the kinematical effects on the line profiles (\cite[Fonfr\'ia et al. 2008]{Fonfria08}).

\begin{figure}[!hbt]
  \centering
  \includegraphics[width=\textwidth]{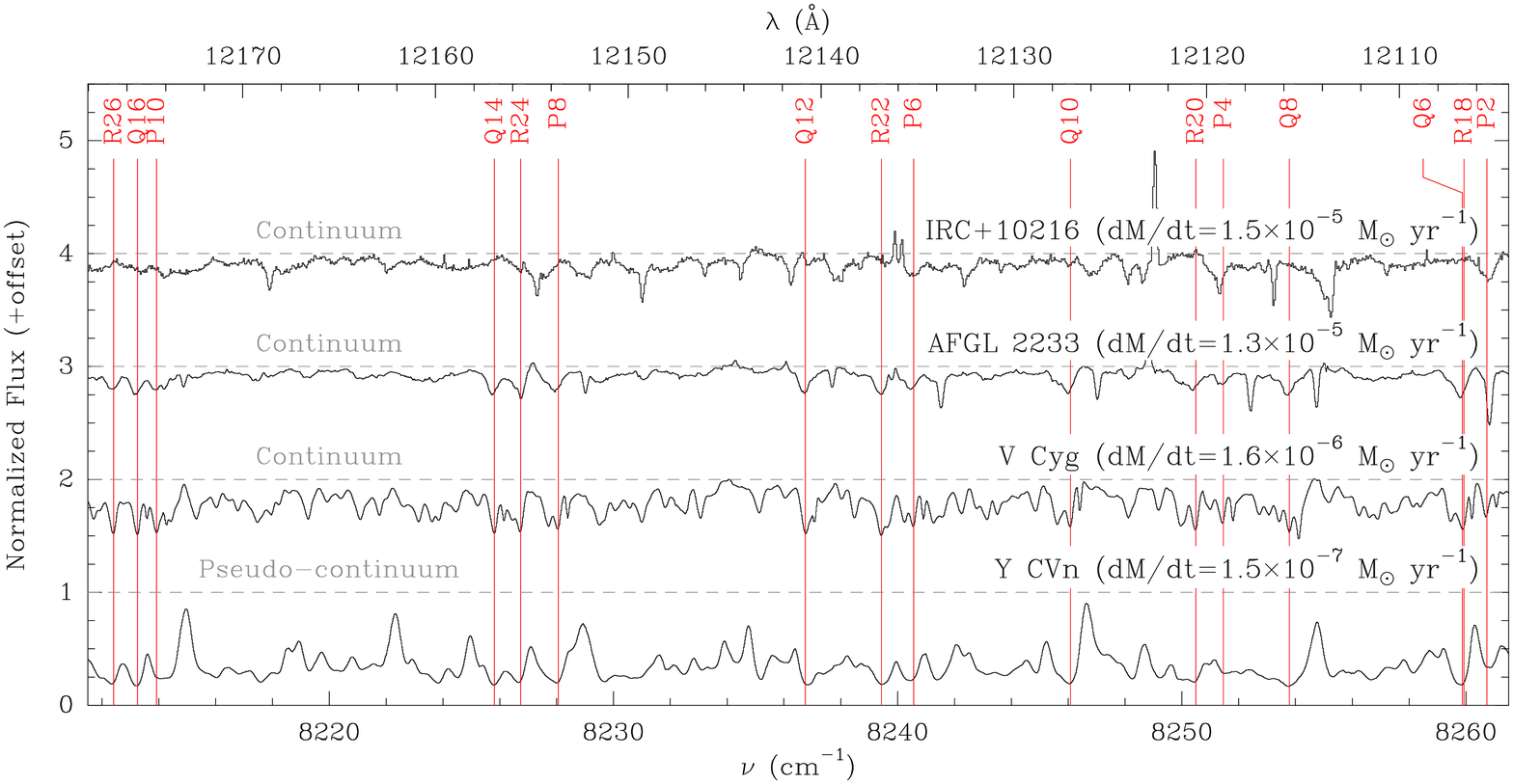}
  \caption{Comparison between the spectra of four C-rich stars with different mass-loss rates in the near-IR acquired with CAHA/CARMENES ($R\simeq 85,000$).
  The C$_2$ lines of the Phillips system $(0,0)$ band are identified.}\label{fig:c2}
  \label{fig:f1}
\end{figure}

The preliminary analysis of our data indicates that the gas-phase C$_2$ column density is different in several AGB stars, as can be seen in Fig.~\ref{fig:c2}.
The C$_2$ bands are clearly stronger in the AGB star with the lowest mass-loss rate (Y CVn) while they are barely noticeable in IRC+10216, which has a mass-loss rate two orders of magnitude higher.
Thus, the higher the mass-loss rate of a C-rich AGB star, the less abundant is C$_2$ in gas-phase around the central object.
Since the mass-loss rate is tightly related to the amount of dust grains available in the innermost envelope, the most plausible explanation of this dependence is that the gas density plays a very important role in the condensation of C$_2$ onto dust grains. The agreggation processes enhanced close to the central star due to the high gas density make of C$_2$ a very good candidate to form the seeds of the dust grains on which other refractory molecules such as SiC$_2$ will condense.

\section{Conclusion}
In this work, we found an evidence of efficient incorporation of SiC$_2$ and C$_2$ onto dust grains, a process that is more efficient at high densities because collisions between particles and coagulation processes become faster. The ring molecule SiC$_2$ emerges as a very likely gas-phase precursor in the process of formation of SiC dust while C$_2$ seems to be highly involved in the building of dust grain seeds in envelopes around C-rich AGB stars. The search for the gas-phase building blocks of dust grains in both carbon- and oxygen-rich AGB stars is a challenging scientific objective, for which astronomical observations still have to provide many answers. Observing other refractory molecules seems mandatory to understand which of the gas-phase molecules that are present in the atmospheres of AGB stars serve as precursor material to feed the process of dust grain formation.

\end{document}